\documentclass[manuscript,a4paper,article]{aastex}
\usepackage{bibentry}
\usepackage{stfloats}
\usepackage{cite}
\usepackage{amsmath}
\usepackage{color}
\usepackage{ulem}
\usepackage{subfigure}
\usepackage{graphicx}
\renewcommand{\raggedright}{\leftskip=0pt \rightskip=0pt plus 0cm}
\raggedright
\newcommand{\Chandra}{{\it Chandra}}
\newcommand{\Suzaku}{{\it Suzaku}}
\newcommand{\XMM}{{\it XMM-Newton}}
\newcommand{\ROSAT}{{\it ROSAT}}
\newcommand{\ASCA}{{\it ASCA}}

\usepackage{epsfig} 
\definecolor{green}{rgb}{0,0.6,0}
\graphicspath{{all/}}
\graphicspath{{extend/}}
\begin{document}

\title{A Chandra Study of {Radial Temperature Profiles} of the Intra-Cluster Medium in 50 Galaxy Clusters}
\author{Zhenghao Zhu\altaffilmark{1}, Haiguang Xu\altaffilmark{1,2}, Jingying Wang\altaffilmark{4}, Junhua Gu\altaffilmark{4}, Weitian Li\altaffilmark{1}, Dan Hu\altaffilmark{1}, Chenhao Zhang\altaffilmark{1}, Liyi Gu\altaffilmark{5}, Tao An\altaffilmark{6}, Chengze Liu\altaffilmark{1}, Zhongli Zhang\altaffilmark{7}, Jie Zhu\altaffilmark{3} \scriptsize AND \normalsize Xiang-Ping Wu\altaffilmark{4}}

\altaffiltext{1}{Department of Physics and Astronomy, Shanghai Jiao Tong University, 800 Dongchuan Road, Minhang, Shanghai 200240, China; email: clsn@sjtu.edu.cn, hgxu@sjtu.edu.cn;}

\altaffiltext{2}{IFSA Collaborative Innovation Center, Shanghai Jiao Tong University, 800 Dongchuan Road, Minhang, Shanghai 200240, China;}

\altaffiltext{3}{Department of Electronic Engineering, Shanghai Jiao Tong University, 800 Dongchuan Road, Minhang, Shanghai 200240, China;}

\altaffiltext{4}{National Astronomical Observatories, Chinese Academy of Sciences, 20A Datun Road, Beijing 100012, China;}

\altaffiltext{5}{SRON Netherlands Institute for Space Research, Sorbonnelaan 2, 3584 CA Utrecht, the Netherlands;}

\altaffiltext{6}{Shanghai Astronomical Observatory, Chinese Academy of Sciences, 80 Nandan Road, Shanghai 200030, China;}

\altaffiltext{7}{Max Planck Institute for Astrophysics, Karl-Schwarzschild-Str. 1, Postfach 1317, D-85741 Garching, Germany.}

\begin{abstract}

In order to investigate the spatial distribution of the ICM temperature in galaxy clusters in a quantitative way and probe the physics behind, we analyze the X-ray spectra of a sample of 50 galaxy clusters, which were observed with the \Chandra\ ACIS instrument in the past 15 years, and measure the radial temperature profiles out to $0.45r_{500}$. We construct a physical model that takes into account the effects of gravitational heating, thermal history (such as radiative cooling, AGN feedback, and thermal conduction) and work done via gas compression, and use it to fit the observed temperature profiles by running Bayesian regressions. The results show that in all cases our model provides an acceptable fit at the 68\% confidence level. To further validate this model we select nine clusters that have been observed with both \Chandra{} (out to $\gtrsim 0.3r_{500}$) and \Suzaku\ (out to $\gtrsim 1.5r_{500}$ ), fit their \Chandra\ spectra with our model, and compare the extrapolation of the best-fits with the \Suzaku\ measurements. We find that the model profiles agree with the \Suzaku\ results very well in seven clusters. In the rest two clusters the difference between the model and observation is possibly caused by local thermal substructures. Our study also implies that for most of the clusters the assumption of hydrostatic equilibrium is safe out to at least $0.5r_{500}$, and the non-gravitational interactions between dark matter and its luminous counterpart is consistent with zero.

\end{abstract}
\section{INTRODUCTION}
 
Galaxy clusters are the most massive virialized systems in our Universe, whose gravitational potential wells are dominated by dark matter that accounts for up to about 90\% of the total mass. Except for the gravity, however, the interaction between dark matter and its luminous counterparts is extraordinary small (e.g., \citealt{munoz15}), thus nearly all of the contemporary studies of clusters have been performed via observing the luminous components, i.e., highly ionized intra-cluster medium (ICM), stellar component, as well as various warm and cool gases, with an emphasis in the X-ray band since the ICM overwhelms other luminous components in mass by a fact of a few (e.g., \citealt{makishima01} and references therein). The temperature of the ICM, which can be directly measured in space-borne observations, is a fundamental quantity that can be used to either characterize the thermodynamic status of the ICM, or calculate the total gravitating mass and various X-ray scaling relations \citep{xu01}, or interpret the Sunyaev-Zeldovich (SZ) effect (\citealt{lin15}). It also provides us with valuable information about astrophysical processes such as AGN feedback, thermal conduction, merger, and radiative cooling. Moreover, the knowledge about the ICM temperature also helps constrain the parameters of our cosmological models (e.g., \citealt{kravtsov12}).

Apparently, in order to carry out physical study of a cluster, an accurate determination of the ICM temperature and its spatial variation within a significant part of the virial radius is crucial. As a standard procedure, after the best-fit gas temperatures obtained for the spectra extracted from a set of adjacent annular or pie regions, both interpolation within each of these regions and extrapolation to larger radii are necessary to obtain a smooth gas temperature profile due to the limited capacity of today's instruments. This invokes a relatively simple and, if possible, universal analytic expression for the temperature profile. As one of the first attempts, \citet{allen01}  analyzed the \Chandra{} pointing observations of seven relaxed clusters, which are located within a redshift range of $0.10 \sim 0.46$, and proposed a 4-parameter empirical profile for the radial distribution of gas temperature.
Using \Chandra{} and \XMM{} data, different but similar temperature profiles were introduced by, e.g., \citet{vikhlinin06} to calculate the total gravitating mass of 13 low-redshift relaxed clusters, \citet{zhang06} to determine the X-ray scaling relations of 14 distant X-ray luminosity, and \citet{ascasibar06} to probe the origin of the cold fronts in the ICM. These profiles and their analogs were evaluated and compared with each other by \citet{gastaldello07}.

Despite the successful applications of the empirical temperature profiles listed above, the following two questions may still be raised: what is the intrinsic physics behind these empirical profiles and how reliable it is when we extrapolate these profiles out to the skirt region of a cluster? In this work, we address this issue by analyzing the X-ray data of a sample of 50 galaxy clusters ($z=0.05\sim 0.46$), which are drawn from \Chandra's 15-year archive, and fitting the observed radial temperature profiles with a physical model that takes into account the effects of gravitational heating during the halo collapse,  thermal history (such as radiative cooling, AGN feedback, and thermal conduction), and work done via gas compression. 
In \S 2 and \S 3, we describe the sample selection criteria and data analysis, respectively. In \S 4, we introduce the model and use it to fit the observed temperature profiles. 
In \S 5, we select nine clusters that have been observed with \Suzaku{} out to or even beyond $1.5r_{500}$ ($r_{500}$ is defined as the radius within which the mean enclosed mass density of the target is 500 times the critical density of the Universe at the target's redshift), extrapolate the best-fit \Chandra{} temperature profiles obtained with our new model to $1.5r_{500}$ and compare the results with the \Suzaku{} measurements. 
We also discuss the applicability of the hydrostatic equilibrium assumption and the constraint on the non-gravitational interaction between dark and luminous matters. In \S 6, we summarize our results. Throughout the work we adopt a flat $\Lambda$CDM cosmology with density parameters $\Omega_{m} = 0.27$ and $\Omega_{\Lambda} = 0.73$, and Hubble constant $\textrm{H}_0 = 71\ \mathrm{km\textbf{ }s}^{-1}\textbf{ }\mathrm{Mpc}^{-1}$. Unless stated otherwise, we used the solar abundance standards of \citet{grevesse98} and quote errors at 68\% confidence level. 
\section{SAMPLE SELECTION AND DATA PREPARATION}
In order to measure and characterize the spatial distribution of gas temperature in a galaxy cluster, both a high signal-to-noise ratio (SNR) and a balance between sufficient angular resolution and complete detector coverage of the target are necessary. To this end, we constructed our \Chandra\ sample by applying the following selection criteria: (1) a large part of the target out to at least $0.45r_{500}$  is fully or nearly fully covered by either the S3 CCD or the I0-3 CCDs of the \Chandra\ Advanced CCD Imaging Spectrometer (ACIS) instrument, (2) the number of photons contained inside $0.45r_{500}$ is no less than 12500 cts, and (3) the target exhibits a relatively regular appearance and possesses no significant substructures. As a result we drew 50 clusters from the \Chandra\ archive, whose redshifts, average temperatures (see \S 3.3), and X-ray luminosities span ranges of $0.05-0.46$, $3-16$ keV, and $7\times 10^{43}-5\times 10^{45}$ erg $\rm s^{-1}$, respectively (Table \ref{sample_tot}).

For each observation, we followed the standard \Chandra\ data processing procedure to prepare the data by using CIAO v4.4 and CALDB v4.4.8, by starting with the ACIS level 1 event files, which were collected with a frame time of 3.2 s telemetered in the FAINT or VFAINT mode. We removed all bad pixels and columns, as well as events with \ASCA\ grades 1, 5, and 7 and carried out corrections for the gain, charge transfer inefficiency (for the observations performed after January 30, 2000), astrometry, and cosmic ray afterglow. By examining the light curves extracted in $0.5-12$ keV from source-free regions or regions less contaminated by the sources on the S3 or I0-3 CCDs, we identified and excluded time intervals contaminated by occasional particle background flares during which the count rate rises to $>120\%$ of the mean value. When available, data of S1 CCD were also analyzed to cross check the determination of the contaminated intervals. All the point sources detected beyond the $3\sigma $ threshold in the ACIS images with CIAO tools \textbf{celldetect} and \textbf{wavedetect} have been masked in the analysis.

\section{DATA ANALYSIS AND RESULTS}
 	\subsection{Background}
 	In order to construct the local background of each observation, we extract the spectrum from one or several separate boundary regions on the S3 CCD or I0-3 CCDs, where the thermal emission of the cluster is relatively weak but usually cannot be neglected, and fit the extracted spectrum { { in $0.4-12$ keV (e.g., \citealt{sun09}; \citealt{vikhlinin05})}}  with a model that consists of an absorbed thermal APEC component (the absorption is fixed to the Galactic value given in \citealt{kalberla05}; \citealt{dickey90}, and the abundance is set to $0.3$ $Z_\odot$ if it is not well constrained; { {e.g., \citealt{panagoulia14}}}), an absorbed power-law component for the Cosmic X-ray Background (CXB; $\Gamma=1.4$; e.g., \citealt{mushotzky00}; \citealt{carter07}), a Galactic emission component (two APEC components with $T_{X}=0.2$ and $0.08$ keV, respectively; e.g., \citealt{gu12}; \citealt{humphrey06}), and a particle-induced hard component calculated from the corresponding \Chandra\ blanksky templates provided by the \Chandra\ Science Center. 
 	The background template employed in this work, i.e., Galactic + CXB + particle components, can be determined after the best-fit is achieved. When available, we also cross checked our background templates with the $0.2-2$ keV \ROSAT\ All-Sky Survey (RASS) diffuse background maps, where the particle component is minor, and obtained consistent results. In order to approximate the field-to-field variation of the Galactic and CXB background components in each observation, in the imaging and spectral analysis that follows we estimate the model parameter errors by taking into account both statistical and systematic uncertainties (10\%; \citealt{kushino02}) in the background. 
 	    
 	\subsection{Radial Temperature Distributions of the ICM}\label{t8a}
 	In order to calculate the azimuthally averaged gas temperature distributions of the sample clusters, for each observation, we extract the \Chandra\ ACIS spectra from $5-7$ concentric annuli, which are all centered on the X-ray peak and cover the $\simeq 0.45r_{500}$. The width of each annulus is determined in such a way that a minimal photon count of 2500 cts in $0.7-7$ keV is guaranteed, while in the outmost annulus the condition that the  photon count is at least twice the background is simultaneously satisfied. We perform the spectral model fittings by employing the X-ray spectral fitting package XSPEC v12.8.2  (\citealt{arnaud96}), and limit the fittings in $0.7-7$ keV to minimize the effect of the instrumental background at higher energies and the calibration uncertainties at lower energies.

 	In the spectral analysis we use the XSPEC model PROJCT to evaluate the influence of the outer spherical shells on the inner ones, and fit the deprojected spectra by applying the optically thin thermal plasma model APEC (\citealt{smith01}), which is absorbed by the foreground photoelectric absorption model WABS. For each cluster, the column density $N_{\rm H}$ of the WABS model is fixed to the corresponding Galactic value (\citealt{kalberla05}; \citealt{dickey90}), and the redshift of the APEC component is fixed to the value given in literature, which can be found in the NASA Extragalactic Database (NED). Whenever the metal abundance of the hot gas is not well constrained, we fix it to 0.3 $Z_\odot$ ({ {e.g., \citealt{panagoulia14}}}). For the innermost annulus, we also attempt to add an additional absorbed APEC component to represent the possible cooler phase gas, the origin of which is often ascribed to the central dominating galaxy (e.g., \citealt{makishima01}). When the F-test shows that the fitting is improved at the 90\% confidence level, we choose to use the two-phase gas model and define the temperature of the intergalactic gas as that of the hot phase. The obtained gas temperature distributions are shown in Figure 1, along with the errors calculated at 68 \% confidences level. 

 	 \subsection{Spatial Distributions of Gas Density, Gas Entropy and Total Gravitational Mass}\label{masscal}
 	We extract the X-ray surface brightness profiles $S_X(R)$ ($R$ is the two dimensional radius) in $0.7-7$ keV from a set of concentric annular bins that are all centered at the X-ray peak of the gas halo, and correct them by applying the exposure maps to remove the effect of vignetting and exposure time fluctuations. The exposure maps are created using the spectral weights calculated for an incident thermal gas spectrum that possesses the same average temperature and metal abundance as the cluster (\S $3.2$). 
 	We assume both hydrodynamic equilibrium and spherical symmetry so that the three-dimensional spatial distribution of the electron density $n_e$ follows either the $\beta$-model, 
 	\begin{equation}
 	n_e(r)=n_0[1+(\frac{r}{r_c})^2]^{-3\beta/2},
 	\end{equation}  
 	or the double-$\beta$ model
 	\begin{equation}
 	n_e(r)=n_{0,1}[1+(\frac{r}{r_{c,1}})^2]^{-3\beta_1/2}+n_{0,2}[1+(\frac{r}{r_{c,2}})^2]^{-3\beta_2/2}
 	\end{equation} 
 	when a significant central surface brightness excess is detected in the inner regions (e.g., \citealt{makishima01}), where $r_c$ and $\beta$ are defined as the core radius and slope parameter, respectively. Given the gas density distribution profiles and the radial distributions of gas temperature and metal abundance obtained by running cubic spline interpolation to the best-fit temperatures and abundances (\S \ref{t8a}), we model the extracted X-ray surface brightness profile as 
 	\begin{equation}\label{s_fit}
 	S_X(R)=\int_{R}^{\infty}\Lambda(T,A)n_en_p(r)\frac{rdr}{\sqrt{r^2-R^2}}+S_{\rm bkg},
 	\end{equation}
 	where $S_{\rm bkg}$ is the diffuse X-ray background, and $\Lambda(T,A)$ is the cooling function. The electron density $n_e(r)$ and gas entropy $K(r)=T(r)n_e(r)^{-2/3}$ are determined when the best-fit to the observed surface brightness profile is achieved by minimizing the $\chi^2$. 
 	Note that the entropy $K(r)$ is the customary ``entropy'' used in the X-ray cluster field, in comparison with the classical definition $S \propto \Delta K/K$ (see \citealt{voit05review} for a review).
 
 	The total gravitating mass of the cluster is calculated as
 	 \begin{equation}
 	 M(<r)=-\frac{r^2k_bT_X}{G\mu m_p}\left[\frac{1}{T_X}\frac{dT_X}{dr}+\frac{1}{n_e}\frac{dn_e}{dr}\right],
 	 \end{equation}
 	 where $\mu$ $=$ $0.61$ is the mean molecular weight per hydrogen atom, $k_b$ is the Boltzmann constant, and $m_p$ is the proton mass. In order to extrapolate the obtained mass profile out to the virial radius, we employ the NFW model \citep{nfw95}   
 	 \begin{equation}
 	  \rho(r)=\frac{\rho_{0}}{\left(1+r/r_s\right)^2r/r_s},
 	 \end{equation}
 	 where $\rho(r)$ is the density of the total gravitational mass, $\rho_0$ and $r_s$ are free parameters in the NFW model. The average temperatures of the clusters are calculated by fitting the spectra extracted in $0.2-0.5$ $r_{500}$, using the same method described in \S 3.2. We show the physical properties of clusters in Table \ref{sample_info}. 
\section{MODELING AND FITTING OF THE OBSERVED TEMPERATURE PROFILES}
\subsection{Effects of Gravity and Non-gravitational Processes}
In this section we attempt to introduce an universal profile to describe the observed gas temperature profiles by taking account the effects of both gravity and non-gravitational processes, the latter  includes the polytropic compression and thermal history (such as radiative cooling, AGN feedback, and thermal conduction) of the gas. For a small test gas element, whose position, mass, and particle number density are $r$, $m^*$, and $n^*$, respectively, we mark the gravitational energy released during the halo collapse as $\Delta E_{\rm G}$, and the energy transferred to it by the non-gravitational processes as $\Delta E_{\rm NG}$. Thus the total energy available to increase the internal energy of the test gas element is expressed as   
\begin{equation}\label{Etot}
E_{\rm total}(r)=\Delta E_{\rm G}(r)+ \Delta E_{\rm NG}(r).
\end{equation}

\noindent {\it Energy Released in the Gravitational Collapse}\\
\indent First let us consider the gravitational energy release of the test gas element during the collapse of the cluster. As in \S \ref{masscal}, we assume that at present time the gravitational potential of the system can be described by the NFW model \citep{nfw95}. Thus when the test gas element falls from infinity to its present position $r$, the gravitational energy release is 
\begin{equation}
\label{Eg}
\Delta E_{\rm G} = \frac{G\int_{0}^{r}\rho(x)4\pi x^2m^*dx}{r}+\int_{r}^{\infty}\frac{G\rho(x)4\pi x^2m^*dx}{x}        =\frac{G\rho_04\pi m^*r_{s}^3}{r}\ln\left(\frac{r+r_{s}}{r_{s}}\right),
\end{equation}
where $G$ is the gravitational constant. This equation denotes the upper limit on the thermal energy increase that can be caused by the gravitational collapse.

\noindent {\it Contributions of the Non-gravitational Processes}\\
 \indent Next we investigate the energy transferred to the test gas element by  non-gravitational processes. As a simple and reasonable assumption, this part of energy is attributed to the work done by gas compression ($\Delta E_{\rm work}$) and the net heating supply determined by the thermal history which typically involves radiative cooling, AGN heating, and thermal conduction ($\Delta E_{\rm heating}$), i.e., $\Delta E_{\rm NG} = \Delta E_{\rm work} + \Delta E_{\rm heating}$. Assuming that ideal gas undergoing polytropic processes that is characterized by the index $n$ during the halo collapse, the state of equations can be written as: 
 \begin{equation}
 p^*V^*=\nu_{\rm mol}^* \mathcal{R}T^* \nonumber {\quad\rm and}
 \end{equation}
 \begin{equation}
 p^*V^{*n}={\rm const},
 \end{equation}
where $p^*$, $V^{*}$, and $T^*$ are the pressure, volume and temperature of the test gas element, respectively, $\mathcal{R}$ is the universal gas constant, and $\nu_{\rm mol}^*=n^*/N_A$ is the mole number as $N_A$ is the Avogadro constant. Thus the work done by the surrounding gas during the compression is calculated by integrating $p^*dV^*$ from the initial state 1 
to the final state 2
\begin{equation}
\Delta E_{\rm work} = \int_{V^*_1}^{V^*_2}-p^*dV^*=p^*_2V_2^{*n}\int_{V^*_2}^{V^*_1}\frac{dV}{V^{*n}}=\frac{p^*_2V^*_2}{n-1}\left[1-\left(\frac{V^*_2}{V^*_1}\right)^{n-1}\right] \nonumber
\end{equation}
\begin{equation}
\label{Ew}
=\frac{1}{n-1}(p^*_2V^*_2-p^*_1V^*_1)=\frac{\nu^*_{\rm mol} \mathcal{R}}{n-1}(T^*_2-T^*_1).
\end{equation}

Now by applying the second law of thermodynamics, we use the entropy change $\Delta S$ to estimate the amount of net non-gravitational heating as $T \Delta S$. Clearly the change of the classical entropy satisfies $\Delta S \propto \Delta K/K$, where $K = T n_e^{-2/3}$ as defined in \S 3.3. Thus we obtain
\begin{equation}\label{Eheating}
\Delta E_{\rm heating}(r) = C_hn^*k_bT(r) (K_{\rm obs}(r) - K_{\rm model}(r)) / K_{\rm obs}(r), 
\end{equation}
where $C_h$ is the scaling factor (\citealt{cnm12}), $K_{\rm obs}$ is the observed entropy profile, and $K_{\rm model}$ is the entropy profile predicted in the hydrodynamical simulation in which only the effect of gravitational energy release is considered. In our calculation, the observed entropy profile is derived by fitting the entropy distribution given in \S 3.3 with an empirical model
\begin{equation}
K_{\rm obs}(r)=K_1+A_1\times r^{\gamma_1},
\end{equation}
where $K_1$, $A_1$ and $\gamma_1$ are parameters constrained by observation. The model predicted entropy profile, on the other hand, is quoted from \citet{voit05} as  
\begin{equation}\label{kmodel}
K_{\rm model}(r)=K_0+A_0\times r^{\gamma_0},
\end{equation}
where $K_0$, $A_0$, and $\gamma_0$ are the parameters given by Eqs. 9, 10 and Figure 5 in \citeauthor{voit05}

\noindent {\it Fraction of Energy Transferred into the Thermal Form}\\
\indent Not all of the energy supply available in the gravitational collapse and non-gravitational processes have been transferred into the thermal energy of the test gas element, and part of them may have been mainly stored as kinetic energy. To account for this energy loss, we introduce an efficiency factor $\eta(r)$ to represent the ratio of the energy transferred into the thermal form to the total energy supplied in both gravitational and non-gravitational processes, i.e., 
\begin{equation}
\label{Tfirst}
\eta(r) = E_{\rm thermal}/E_{\rm total} = \frac{3}{2}n^*k_bT(r) / E_{\rm total}
\end{equation}
where $E_{\rm total}$ is determined by Eqs. \ref{Etot}, \ref{Eg}, \ref{Ew} and \ref{Eheating}. Since  
$E_{\rm thermal}/E_{\rm total} = 
\bar v_{\rm thermal}^2/(\bar v_{\rm thermal}^2  + \bar v_{\rm kinetic}^2) = 
P_{\rm thermal}/P_{\rm total}=
1 - P_{\rm kinetic}/P_{\rm total}$,
where $P_{\rm thermal}$, $P_{\rm kinetic}$, and $P_{\rm total}$ are the thermal, kinetic, and total pressure, respectively. Using $P_{\rm kinetic} / P_{\rm total}$ 
given in \citet{battaglia2011} we calculate the efficiency as
\begin{equation}
\label{eta}
\eta(r) = 1 - \alpha_0(1+z)^{\beta_h}\left(\frac{r}{R_{500}}\right)^{n_{\rm nt}}\left(\frac{M_{200}}{3\times 10^{14}M_{\rm sun}}\right)^{n_M}, 
\end{equation}  
where $\alpha_0=0.18\pm 0.06$, $\beta_h=0.5$, $n_{\rm nt}=0.8 \pm 0.25$, and  $n_M \simeq 0.2$ as calculated in \citet{battaglia2011}.

\noindent{\it Gas Temperature Profile}\\
\indent Substituting Eqs. \ref{Etot}, \ref{Eg}, \ref{Ew}, \ref{Eheating}, and \ref{eta} into \ref{Tfirst} , and noting that $T^*_1$ is the effective initial temperature defined under the  quasi-static assumption, which is estimated to be of the order  $T_G=2\Delta E_G/(3n^*k_b)$ and $T^*_2=T(r)$ is the observed temperature today, we obtain 
\begin{footnotesize}
\begin{equation}
T(r)=\frac{2\eta(r)}{3n^*k_b}\left(C_hn^*k_bT(r)\frac{K_{\rm obs}(r)-K_{\rm model}(r)}{K_{\rm obs}(r)}+\frac{\nu^*_{\rm mol} \mathcal{R}}{n-1}(T(r)-T_G)+\frac{G\rho_04\pi m^*r_{s}^3}{r}\ln\left(\frac{r+r_{s}}{r_{s}}\right)+n^*e_0\right) 
\end{equation}
\end{footnotesize}where $e_0$ is the average initial energy for a single particle in the gas element, which should be approximately zero. Rewriting this immediate yields the temperature profile 
\begin{equation}\label{model}
T(r)=\frac{T_0+(1-C_3)C_1\rho_0r_s^3\ln((r_s+r)/r_s)/r}{1/\eta(r)-C_3-C_2(K_{\rm obs}(r)-K_{\rm model}(r))/K_{\rm obs}(r)},
\end{equation}
where the temperatures are measured in keV.
In this formula, $T_{0}=2e_0/3k_b$ is the initial gas temperature and is assumed to be $\simeq 0$. The parameter $C_1=8\pi G\mu m_p/3$, where $\mu m_p=m^*/n^* \simeq 0.61m_p$ is a fixed combination of physical constants. $C_2=2C_h/3$ is related to the scaling factor $C_h$ (c.f., Eq. \ref{Eheating}), which equals $1.5$ in an isochoric process (e.g., \citealt{cnm12}). $C_3=2/(3(n-1))$, where $n$ is the polytropic index to be decided in the fitting. The parameters $\eta(r)$ (c.f., Eq. \ref{eta}), $K_{\rm model}(r)$ (c.f., Eq. \ref{kmodel}), $K_{\rm obs}$, $\rho_0$ and $rs$ as well as their error ranges are constrained by observations, simulations and/or scaling relations in the Bayesian model fittings (\S \ref{fitting}).  
  \subsection{Fitting Method and Results}\label{fitting}
  
  By assuming that the parameters in our model are independent, which can be considered as a good approximation in practice, we use Eq. \ref{model} to fit the gas temperature distributions observed with \Chandra{} (\S 3.2) by running Bayesian regressions (e.g., \citealt{andreon13}; \citealt{andreon12}). Compared with chi-squared test and maximum-likelihood estimation, this approach has the advantage that it incorporates observation errors in the model meanwhile it can quantify both the intrinsic scatter and the uncertainties of the known model parameters. To be specific, we apply Bayes theorem to express the posterior probability distribution as the product of the prior distribution, which is tightly constrained by the uncertainty ranges of known model parameters (c.f., Eq. \ref{model}), and the likelihood function determined according to the error ranges of gas temperature allowed by the observation, i.e.,
  \begin{align}
  {\rm Pr(fitted\; parameter | observed\; data)} \nonumber
  &\propto 
  {\rm Pr( fitted\; parameter) \times} \\ 
   &{\rm Pr( observed\; data | fitted\; parameter)}
  \end{align}
  and maximize it with Powell's method \citep{powell64}, which is powerful for calculating the local maximum of a continuous but complex function. When the maximum is found, the best-fit temperature profile is achieved. We plot the best-fit profiles in Figure 1, where the 68\% error bands of the model computed by 
  Monte-Carlo simulation are shown in shadow. {{To evaluate the goodness of the fittings, we introduce the model efficiency $R_{\rm eff}$ (See \citealt{nash70} and \citealt{engeland02}), which is defined as 
  \begin{equation}
  R_{\rm eff}=\frac{1}{N_{\rm bin}}\sum_{i=1}^{N_{\rm bin}}R_{\rm eff,i}, \quad {\rm and} \nonumber
  \end{equation}
  \begin{equation}
  R_{\rm eff,i}=1-\frac{\sum_{n=1}^{N_{\rm sim}}\left(T_{i,n,\rm sim}-T_{i,n,\rm mod}\right)^2}{\sum_{n=1}^{N_{\rm sim}}\left(T_{i,n,\rm sim}-T_{i,n,\rm obs}\right)^2},
  \end{equation}
  where $N_{\rm bin}$ is the bin number of the observed temperature distribution, $N_{\rm sim}=1000$ is the total number of Monte-Carlo simulation, and $T_{\rm sim}$, $T_{\rm mod}$ and $T_{\rm obs}$ represent the simulated, model-predicted and observed temperatures, respectively. As shown in Table \ref{sample_info}, for all of the clusters, the new profile proposed in this work gives an acceptable fit ($R_{\rm eff} \sim 0$), although in five cases ($R_{\rm eff} \sim -1$) one data bin deviates the from model prediction slightly (68 \% confidence level).
  To describe the degree of the violation to the hydrostatic equilibrium, we also plot the best-fit non-thermal energy fraction $1-\eta(r)$ as a function of radius for all clusters in Figure 2.  }}

\section{DISCUSSION}

\subsection{A Comparison with {\it Suzaku}'s Results }
In order to further validate our temperature profile model we select nine galaxy clusters (Table \ref{sample_add}), seven of which are not involved in our sample due to either a small ($< 0.45r_{500}$) field coverage or a low signal-to-noise ratio in the \Chandra{} observation, calculate their gas temperature profiles with the ACIS S3 or I0-3 data, extrapolate the best-fit model profiles derived with Eq. \ref{model} to $1.5r_{500}$, and compare the results with those measured with the \Suzaku\ satellite (see references in Table \ref{sample_add}), as shown in Figure 3. The difference between the temperatures measured with \Chandra{} and \Suzaku{} caused by the energy dependence of the stacked residuals ratios, (i.e., the energy-dependent difference between the effective areas of the two instruments even after careful calibrations; \citealt{schellenberger15}; \citealt{kettula13}; \citealt{nevalainen10}), has been compensated by using the data provided by \citet{schellenberger15}\footnote{The authors listed the \Chandra\ and \XMM{} measurements of gas temperature to show their difference, meanwhile it is known that the \Suzaku\ and \XMM{} measurements of temperature are consistent with each other in $0.7-7$ keV.}.

We find that within $1.5r_{500}$ (roughly $1.0r_{200}$) of seven clusters our model profiles are in good agreement with the \Suzaku{} measurements at the 68\% confidence level. For A1835 the model prediction is higher than the \Suzaku\ measurement in $r \gtrsim r_{500}$, where a break on the observed gas entropy distribution is detected against the theoretical power-law distribution. This is an indication of the appearance of additional cooler gas, which can be  associated with contacting regions between the cluster and large-scale structure environment \citep{ichikawa13}.
For A2029 our model underestimates the \Suzaku\ temperatures in $0.13-0.5$ $r_{500}$, which might be due to substructures of the cluster.

\subsection{Are the Sample Clusters in Hydrodynamic Equilibrium?}
Hydrostatic equilibrium has been widely assumed in X-ray imaging spectroscopic analysis of galaxy clusters because, e.g., by adopting this assumption, the projection effect can be well restricted in measuring the X-ray mass, meanwhile the scatter of X-ray mass measurements is about a factor of two smaller than that of the lensing masses. However, both observations and simulations show that the X-ray mass measurements are generally biased low by $5-20\%$ due to violation of the hydrostatic equilibrium due to mergers and bulk motions of the gas (e.g., \citealt{meneghetti10}), especially in the skirt regions. When this violation occurs, we may employ the efficiency factor $\eta(r)$ introduced in \S4.1 to describe the degree of the violation as 
$E_{\rm non\text{-}thermal} / E_{\rm total} = 1-\eta(r)$,
which is typical $ <10\%$ inside $0.3r_{200}$ and increases to $\simeq 20 \%$ at $r_{200}$ \citep{battaglia2011}. 
In our model fittings of the \Chandra\ temperature profiles of the sample clusters (\S4.2), we usually obtain 
$E_{\rm non\text{-}thermal} / E_{\rm total} \lesssim 0.12$ 
at small and intermediate radii, whereas $E_{\rm non\text{-}thermal} / E_{\rm total}$ reaches to $\simeq 0.25$ at $r_{500}$ (Figure 2).

If we force $E_{\rm non\text{-}thermal} / E_{\rm total} \equiv 0$ throughout the cluster, we will obtain gas temperatures much higher than the observation in $\gtrsim r_{500}$ in some cases. As an example in Figure 3 we show the model predicted temperature profiles for the nine clusters discussed in \S5.1 (light blue solid lines) when $E_{\rm non\text{-}thermal} / E_{\rm total} \equiv 0$ is set. In three clusters (A1835, A2142, and PKS 0745-191) the model profiles significantly deviate from the observed data, indicating that in these clusters the hydrostatic equilibrium is not well established in $ 0.45-1.5r_{500}$. For the rest six clusters, however, such a deviation is not seen in $r \le r_{500}$ or even in $r \le r_{200}$ within the error ranges. 
These results imply that for these clusters the assumption of hydrostatic equilibrium is safe at small and intermediate radii.

\subsection{None Gravitational Interaction between Dark Matter and Baryonic Matter }
The non-gravitational interactions between dark matter and luminous matter are expected to be very weak, therefore the kinetic energy of the dark matter can hardly be transferred to the hot gas. To test this we modify our model by adding a new free parameter $f_{\rm DMenergy}$ to describe the percentage of the dark matter's kinetic energy that has been transfered to the gas. Assuming that the baryonic fraction of the total gravitating mass is 16\%, the model temperature profile Eq. \ref{model} is modified to be
\begin{equation}
T(r)=\frac{T_0+(1+(1/0.16-1)f_{\rm DMenergy} - C_3)C_1\rho_0r_s^3\ln((r_s+r)/r_s)/r}{1/\eta(r)-C_3-C_2(K_{\rm obs}(r)-K_{\rm model}(r))/K_{\rm obs}(r)}.
\end{equation} 

We find that $f_{\rm DMenergy}$ given by the fittings are consistent with zero for all clusters within the computational accuracy. This is accordant with the measurement results by astronomical simulation \citep{munoz15} and particle physics experiments \citep{aprile12}, which in turn verify the model.

\section{SUMMARY}
We investigate the gas temperature profiles in a sample of 50 clusters observed with \Chandra\ by introducing a new temperature profile model, which takes into account the effects of both the gravitational heating, thermal history (such as radiative cooling, AGN feedback, and thermal conduction) and work done via gas compression, and use it to fit the observed temperature profiles by running Bayesian regressions. In all cases our model can provide an acceptable fit at the 68\% confidence level. Also we find that by extrapolating the best-fit \Chandra\ temperature profiles derived with our model to $1.5r_{500}$ the results agree very well with the \Suzaku\ measurements. With the new model we show that for most clusters assumption of hydrostatic equilibrium is safe out to at least $0.5r_{500}$.

This work was supported by the Ministry of Science and Technology of China
(grant No. 2013CB837900),
the National Science Foundation of China
(grant Nos. 11125313, 11203017, 11433002, 61271349, and 61371147),
the Chinese Academy of Sciences
(grant No. KJZD-EW-T01),
Science and Technology Commission of Shanghai Municipality
(grant No. 11DZ2260700),
and Shanghai Key Lab for Particle Physics and Cosmology (SKLPPC) (Grant No. 11DZ2260700).

\appendix

\begin{deluxetable}{lllll}
\tablecaption{Basic information of sample members\label{sample_tot}\tablenotemark{a}}
\tabletypesize{\scriptsize}
\tablewidth{0pt}
\tablehead{\colhead{Name} & \colhead{Obsid} &\colhead{RA} &\colhead{DEC}  &\colhead{Redshift}}
\startdata
2PIGGJ0011.5-2850 & 5797 & 00:11:21.618 & -28:51:21.47 & 0.0625 \\
3C295 & 2254 & 14:11:20.676 & +52:12:08.99 & 0.4641\\
3C388 & 5295 & 18:44:02.014 & +45:33:30.68 & 0.0917\\
A1132 & 13376    & 10:58:26.518 & +56:47:35.34 & 0.1363\\
A115N & 13458 & 00:55:50.691 & +26:24:37.25 & 0.1971\\
A115S & 13458 & 00:55:59.283 & +26:19:56.58 & 0.1971 \\
A1201 & 9616  & 11:12:54.450 & +13:26:00.44 & 0.1688\\
A1423 & 11724  & 11:57:17.344 & +33:36:40.75 & 0.213\\
A1553 & 12254  & 12:30:46.941 & +10:33:17.65 & 0.1652\\
A1664 & 7901  & 13:03:42.370 & -24:14:43.66 & 0.1283\\
A1763 & 3591  & 13:35:18.357 & +40:59:58.65 & 0.223\\
A1835 & 6880  & 14:01:01.971 & +02:52:40.88 & 0.2532\\
A2061 & 10449 & 15:21:10.711 & +30:37:58.27 & 0.0784\\
A2163 & 1653  & 16:15:46.098 & -06:08:55.13 & 0.203\\
A2204 & 7940  & 16:32:46.981 & +05:34:31.87 & 0.1522\\
A2218 & 1666  & 16:35:51.863 & +66:12:37.99 & 0.1756\\
A2219 & 896   & 16:40:20.250 & +46:42:30.65 & 0.2256\\
A2244 & 4179  & 17:02:42.316 & +34:03:34.27 & 0.0968\\
A2249 & 12284 & 17:09:44.399 & +34:27:24.18 & 0.0816\\
A2255 & 894   & 17:12:44.779 & +64:04:29.86 & 0.0806\\
A2384 & 4202  & 21:52:21.450 & -19:32:54.93 & 0.0943\\
A2409 & 3247  & 22:00:52.915 & +20:58:27.42 & 0.1479\\
A2420 & 8271  & 22:10:19.165 & -12:10:17.82 & 0.0846\\
A2457 & 12276 & 22:35:41.641 & +01:29:11.35 & 0.0594\\
A2537 & 9372  & 23:08:22.105 & -02:11:29.34 & 0.295\\
A2667 & 2214  & 23:51:39.337 & -26:05:03.22 & 0.23\\
A3088 & 9414  & 03:07:01.858 & -28:39:55.57 & 0.2534\\
A3158 & 3712  & 03:42:51.735 & -53:37:48.13 & 0.0597\\
A3528 & 8268  & 12:54:40.759 & -29:13:40.15 & 0.0530\\
A3695 & 12274 & 20:34:47.434 & -35:49:03.38 & 0.0894\\
A3827 & 7920  & 22:01:53.464 & -59:56:46.05 & 0.0984\\
A3921 & 4973  & 22:49:57.612 & -64:23:43.40 & 0.0928\\
A399 & 3230   & 02:57:51.172 & +13:02:37.12 & 0.0718\\
A520 & 4215   & 04:54:09.806 & +02:55:23.41 & 0.199\\
A644 & 2211   & 08:17:25.497 & -07:30:39.40 & 0.0704\\
A665 & 13201  & 08:30:59.962 & +65:50:35.49 & 0.1819\\
A773 & 5006   & 09:17:52.853 & +51:43:39.91 & 0.217\\
MACSJ0035.4-2015 & 3262 & 00:35:26.339 & -20:15:47.37 & 0.364\\
MACSJ1206.2-0847 & 3277 & 12:06:12.482 & -08:48:05.73 & 0.44\\
MS0906.5+1110 & 924      & 09:09:12.615 & +10:58:32.37 & 0.18 \\
RXCJ0528.9-3927 & 4994   & 05:28:52.801 & -39:28:21.11 & 0.2839\\
RXCJ0605.8-3518 & 15315  & 06:05:53.977 & -35:18:08.60 & 0.141\\
RXCJ0638.7-5358 & 9420     & 06:38:47.101 & -53:58:28.78 & 0.2216\\
RXCJ1329.7-3136 & 4165        & 13:29:47.398 & -31:36:19.58 & 0.0495\\
RXCJ1504.1-0248 & 5793   & 15:04:07.630 & -02:48:15.95 & 0.2153\\
RXCJ2218.6-3853 & 15101  & 22:18:39.634 & -38:53:56.35 & 0.1379\\
RXJ0006.3+1052 & 12251 & 00:06:20.557 & +10:51:52.98 & 0.1675\\
RXJ0439.0+0715 & 3583    & 04:39:00.678 & +07:16:03.98 & 0.23 \\
RXJ1023.7+0411 & 909     & 10:23:39.648 & +04:11:11.90 & 0.2906\\
RXJ1720.1+2638 & 4361    & 17:20:10.115 & +26:37:29.61 & 0.164\\
\enddata
\tablenotetext{a}{Right ascension and declination are in J2000.0.}
\end{deluxetable}

\begin{deluxetable}{lllllr}
\tablecaption{Average gas temperature, $r_{500}$, total gravitating mass and {0.5-10} keV luminosity within $r_{500}$, {{and model efficiency}} \label{sample_info} }
\tabletypesize{\scriptsize}
\tablewidth{0pt}
\tablehead{\colhead{Name} & \colhead{Average temperature\tablenotemark{a}} &\colhead{$M_{500}$} &\colhead{$r_{500}$ } &\colhead{ $L_{X,500}$($0.5-10$ keV)} & {{$ R_{\rm eff}$}}\\
 &\qquad\qquad(keV) & ($10^{14}$ $M_\odot$) & \;\, (kpc) & \quad \;\,($10^{44}$ erg $\rm s^{-1}$) &  }

\startdata
2PIGGJ0011.5-2850 & $4.04_{-0.18}^{+0.18}$ & $2.34_{-0.51}^{+0.91}$ & $913_{-72}^{+106}$& $1.93 \pm 0.19$ & $-0.37$\\
3C295 & $6.98_{-0.72}^{+0.90}$ & $3.48_{-0.52}^{+0.74}$ & $910_{-48}^{+60}$& $11.0 \pm 0.7$ & $-0.17$ \\
3C388 & $3.08_{-0.21}^{+0.11} $ & $1.06_{-0.20}^{+0.39} $&$ 694_{-46}^{+76}$ & $0.70 \pm 0.08$ & $-0.33$ \\
A1132 & $10.2_{-1.1}^{+1.2}$ & $ 11.0_{-2.2}^{+2.7} $ & $1494_{-107}^{+114} $ & $11.3 \pm 0.9 $ & $0.33$ \\
A115N & $10.2_{-0.5}^{+0.5}$ & $6.36_{-1.36}^{+1.31}$ & $1221_{-94}^{+79}$ & $8.91 \pm 0.52$ & $-0.77$\\
A115S & $ 8.19_{-0.3}^{+0.31}$ &$5.78_{-1.36}^{+1.56}$ & $1183_{-101}^{+98}$ & $8.17 \pm 0.47$ & $0.18$ \\
A1201 & $ 6.45_{-0.31}^{+0.31}$ & $4.68_{-1.64}^{+2.47} $ & $1112_{-149}^{+169}$ & $5.95 \pm 0.70$ & $-0.30$\\
A1423 & $7.48_{-0.70}^{+0.75}$ &$2.03_{-0.58}^{+0.60}$ & $866_{-92}^{+78}$ & $1.12 \pm 0.15$ & $-0.37$ \\
A1553 & $7.74_{-0.71}^{+0.71}$ & $7.97_{-1.26}^{+1.96}$ & $1330_{-74}^{+101}$ & $7.35 \pm 0.53$ & $-0.16$  \\
A1664 & $5.35_{-0.20}^{+0.26}$ & $3.79_{-0.38}^{+0.40}$ & $1050_{-38}^{+36}$ & $4.84 \pm 0.32$ & $-0.63$\\
A1763 & $9.42_{-0.82}^{+0.82}$ & $5.18_{-0.80}^{+1.23}$ & $1131_{-61}^{+83}$ & $14.2 \pm 0.9$ & $-0.03$ \\
A1835 & $11.5_{-0.6}^{+0.6}$ & $10.8_{-0.9}^{+1.1}$ & $1430_{-41}^{+47}$ & $37.7 \pm 2.2$ & $-0.97$ \\
A2061 & $5.05_{-0.17}^{+0.17}$ & $3.63_{-0.51}^{+1.73}$ & $1051_{-51}^{+146} $ & $2.32 \pm 0.21$ & $0.45$\\
A2163 & $16.1_{-0.5}^{+0.5}$ & $19.5_{-2.5}^{+2.1}$ & $1770_{-79}^{+61} $ & $49.8 \pm 2.8$  & $0.19$\\
A2204 & $9.52_{-0.31}^{+0.31}$ & $8.62_{-0.41}^{+0.48}$ & $1371_{-22}^{+25} $ & $27.3 \pm 1.5$ & $-0.42$\\
A2218 & $7.11_{-0.37}^{+0.38}$ & $6.50_{-0.31}^{+0.39}$ & $1239_{-20}^{+24}$ & $9.52 \pm 0.57$ & $-0.07$ \\
A2219 & $12.7_{-0.7}^{+0.6} $ & $21.3_{-4.3}^{+3.4}$ & $1810_{-131}^{+93}$ & $34.1 \pm 2.0$ & $0.08$ \\
A2244 & $6.15_{-0.14}^{+0.14} $ & $4.12_{-0.6}^{+0.4}$ & $1090_{-54}^{+34}$ & $8.51 \pm 0.51$ & $-0.88$\\
A2249 & $6.98_{-0.44}^{+0.66} $ & $4.03_{-1.14}^{+5.17}$ & $1087_{-114}^{+344}$ & $3.22 \pm 0.51$ & $-0.12$ \\
A2255 & $6.64_{-0.14}^{+0.14} $ & $4.23_{-0.48}^{+0.81}$ & $1105_{-44}^{+66}$ & $4.50 \pm 1.28 $ & $-1.23$\\
A2384 & $ 5.85_{-0.27}^{+0.27}$ & $2.77_{-0.35}^{+0.26} $ & $956_{-42}^{+29} $ & $2.50 \pm 0.19$ & $-0.37$ \\
A2409 & $6.52_{-0.45}^{+0.45} $ & $6.20_{-1.14}^{+1.68}$ & $1230_{-81}^{+102}$ & $9.52 \pm 0.65$ & $-0.32$ \\
A2420 & $4.36_{-0.29}^{+0.38} $ & $3.73_{-0.83}^{+1.55}$ & $1058_{-86}^{+130}$ & $4.86 \pm 0.51$ & $0.20$ \\
A2457 & $3.92_{-0.27}^{+0.26} $ &　$2.76_{-1.05}^{+2.12}$ & $965_{-150}^{+193}$ & $1.05 \pm 0.08$& $0.02$ \\
A2537 & $8.84_{-0.97}^{+1.24} $ & $4.30_{-0.37}^{+0.92}$ & $1036_{-32}^{+67}$ & $9.94 \pm 0.64$ & $0.04$ \\
A2667 & $7.97_{-0.74}^{+0.89} $ & $6.51_{-1.74}^{+2.52}$  & $1217_{-120}^{+140}$ & $24.9 \pm 1.6$ & $-0.33$ \\
A3088 & $ 8.05_{-0.78}^{+1.12} $ & $5.49_{-0.81}^{+2.06} $ & $1141_{-59}^{+128} $ & $13.4 \pm 1.0$ & $0.42$\\
A3158 & $4.96_{-0.88}^{+0.88} $ & $3.22_{-0.37}^{+0.50}$ & $1016_{-41}^{+49} $ & $4.66 \pm 0.42$  & $-0.23$\\
A3528 & $5.03_{-0.34}^{+0.34}$ & $2.18_{-0.41}^{+0.62}$ & $893_{-61}^{+75} $ & $1.61 \pm 0.15$ & $0.12$\\
A3695 & $6.47_{-0.41}^{+0.41}$ & $4.10_{-1.84}^{+2.40}$ & $1091_{-197}^{+181}$ & $5.05 \pm 0.55$ & $0.14$\\
A3827 & $7.61_{-0.21}^{+0.21}$ &$4.53_{-0.28}^{+0.34}$ & $1124_{-23}^{+28}$ & $9.76\pm 0.63$ & $0.03$\\
A3921 & $5.92_{-0.22}^{+0.22}$ & $4.02_{-0.37}^{+0.73}$ & $1083_{-34}^{+62}$ & $4.57 \pm 0.31$ & $-0.63$\\
A399 & $8.18_{-0.25}^{+0.25}$ &　$4.67_{-0.79}^{+1.46} $ & $1145_{-72}^{+108}$ & $5.76 \pm 0.37$ & $0.13$\\
A520 & $9.04_{-0.80}^{+0.80}$ & $6.78_{-0.48}^{+0.74} $ & $1246_{-30}^{+44}$ & $13.2 \pm 0.72 $ & $-0.10$ \\
A644 & $8.28_{-0.21}^{+0.21}$ & $6.70_{-0.58}^{+0.37} $ & $1292_{-39}^{+23}$ & $9.10 \pm 0.59 $ & $-0.06$ \\
A665 & $12.2_{-0.8}^{+0.7}$ & $7.65_{-2.25}^{+1.90} $ & $1305_{-143}^{+100} $ & $16.1 \pm 1.0$ & $-0.26$ \\
A773 & $7.67_{-0.59}^{+0.60}$ & $5.26_{-1.20}^{+3.72} $ & $1139_{-95}^{+222} $ & $11.0 \pm 0.9$  & $0.33$\\
MACSJ0035.4-2015 & $6.45_{-0.46}^{+0.55}$ & $5.41_{-1.01}^{+0.81}$ & $1093_{-72}^{+52} $& $22.2 \pm 1.6$ & $0.36$\\
MACSJ1206.2-0847 & $15.8_{-2.1}^{+2.5}$ & $11.7_{-2.9}^{+3.7}$ & $1374_{-123}^{+132}$ & $41.5 \pm 2.4$ & $0.39$\\
MS0906.5+1110 & $5.46_{-0.35}^{+0.42}$ & $4.09_{-0.30}^{+0.48}$ & $1059_{-27}^{+53}$ & $5.98 \pm 0.39$ & $-0.06$\\
RXCJ0528.9-3927 & $9.08_{-0.90}^{+1.18}$ & $8.18_{-3.79}^{+2.48}$ & $1290_{-241}^{+119}$& $19.4 \pm 1.5$ & $0.17$\\
RXCJ0605.8-3518 & $6.64_{-0.50}^{+0.64}$ & $3.91_{-0.64}^{+1.78}$ & $1057_{-61}^{+141} $&$ 8.55\pm 0.74$ & $-0.35$\\
RXCJ0638.7-5358 & $9.73_{-0.81}^{+0.81}$ & $7.53_{-2.02}^{+3.27}$ & $1281_{-126}^{+164}$ & $ 23.3 \pm 1.4 $ & $0.02$\\
RXCJ1329.7-3136 & $3.45_{-0.14}^{+0.17}$ & $0.79_{-0.12}^{+0.22}$ & $637_{-35}^{+56}$ & $0.88 \pm 0.06$ & $-1.27$ \\
RXCJ1504.1-0248 & $11.0_{-0.8}^{+1.1}$ & $9.29_{-0.48}^{+0.62}$ & $1378_{-25}^{+30}$ & $47.4\pm 2.7$ & $-0.16$ \\
RXCJ2218.6-3853 & $8.67_{-0.9}^{+1.38}$ & $9.36_{-2.00}^{+4.37}$ &　$1415_{-109}^{+193}$ & $9.15 \pm 0.54$ & $0.40$ \\
RXJ0006.3+1052 & $6.85_{-0.36}^{+0.53}$ & $2.83_{-0.41}^{+0.28}$ & $941_{-48}^{+30}$ & $4.87 \pm 0.41$ & $-1.28$ \\
RXJ0439.0+0715 & $6.51_{-0.42}^{+0.52}$ & $4.34_{-0.61}^{+1.15}$ & $1064_{-52}^{+87}$ & $12.9 \pm 0.9$ & $-0.40$ \\
RXJ1023.7+0411 & $9.26_{-0.61}^{+0.62}$ & $6.64_{-0.55}^{+0.97}$ & $1200_{-34}^{+56}$ & $35.4 \pm 2.0$ & $-0.04$\\

RXJ1720.1+2638 & $8.38_{-0.49}^{+0.62}$ & $6.22_{-0.76}^{+0.62}$ & $1225_{-57}^{+38}$ & $ 15.5 \pm 0.95$ & $-0.02$\\
\enddata
\tablenotetext{a}{Average temperature is calculated for $0.2-0.5r_{500}$ region.}

\end{deluxetable}

\begin{deluxetable}{lllllll}
\tablecaption{Nine clusters selected to compare our model predictions with the \Suzaku\ measurements\tablenotemark{a}  \label{sample_add}}
\tabletypesize{\scriptsize}
\tablewidth{0pt}
\tablehead{\colhead{Name} & \colhead{Obsid} & \colhead{RA} & \colhead{DEC}& \colhead{Redshift} & \colhead{Reference\tablenotemark{b}} }

\startdata
A1413 & 5003  & 11:55:18.168 & +23:24:24.49 & 0.1427  & \citealt{hoshino10}\\ 
A1689 & 6930  & 13:11:29.472 & -01:20:30.12 & 0.183   & \citealt{kawaharada10}\\ 
A1795 & 10898 & 13:48:52.742 & +26:35:27.63 & 0.0625  & \citealt{bautz09}\\ 
A1835 & 6880  & 14:01:01.971 & +02:52:40.88 & 0.2532  & \citealt{ichikawa13}\\ 
A2029 & 4977  & 15:10:56.195 & +05:44:43.36 & 0.07728 & \citealt{walker12a} & \\ 
A2142 & 5005  & 15:58:19.972 & +27:13:58.50 & 0.0909  & \citealt{akamatsu11}\\ 
A2204 & 7940  & 16:32:46.981 & +05:34:31.87 & 0.1522  & \citealt{reiprich09}\\ 
A780 & 575 & 09:18:05.352 & -12:05:46.55 & 0.0539  & \citealt{sato12}\\
PKS 0745-191 & 6103 & 07:47:31.430 & -19:17:42.29 & 0.1028 & {\citealt{walker12b}}\\
\enddata
\tablenotetext{a}{Right ascension and declination are in J2000.0.}
\tablenotetext{b}{90\% errors quoted from the references, except for A2029 and RXJ0747.5-1917, for which the errors are quote at the $68\%$ confidence level. The projection effect are not corrected expect for A2029 { {and PKS 0745-191}}.}

\end{deluxetable}
\begin{figure}

\includegraphics[width=1.0\textwidth,height=0.9\textwidth]{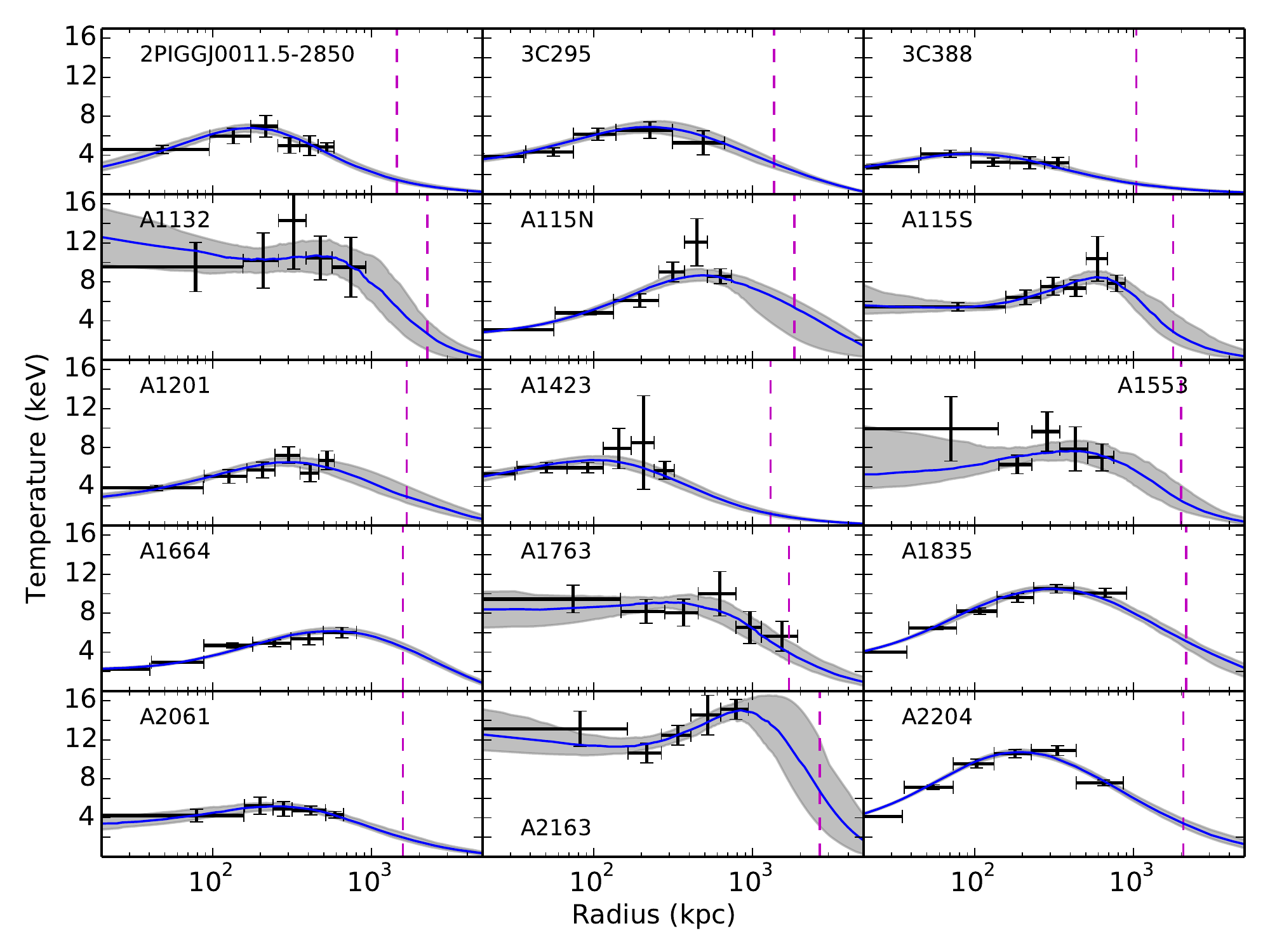}
\caption{Best-fit gas temperature profiles obtained with our model (dark blue curves), along with the 68\% errors (shadow), to the \Chandra\ measurements (black crosses) { {after the projection effect is corrected (\S \ref{t8a}). The vertical dashed lines indicate $1.5r_{500}$ of each cluster. }}}

\end{figure}
\setcounter{figure}{0}
\begin{figure}

\includegraphics[width=1.0\textwidth,height=1.0\textwidth]{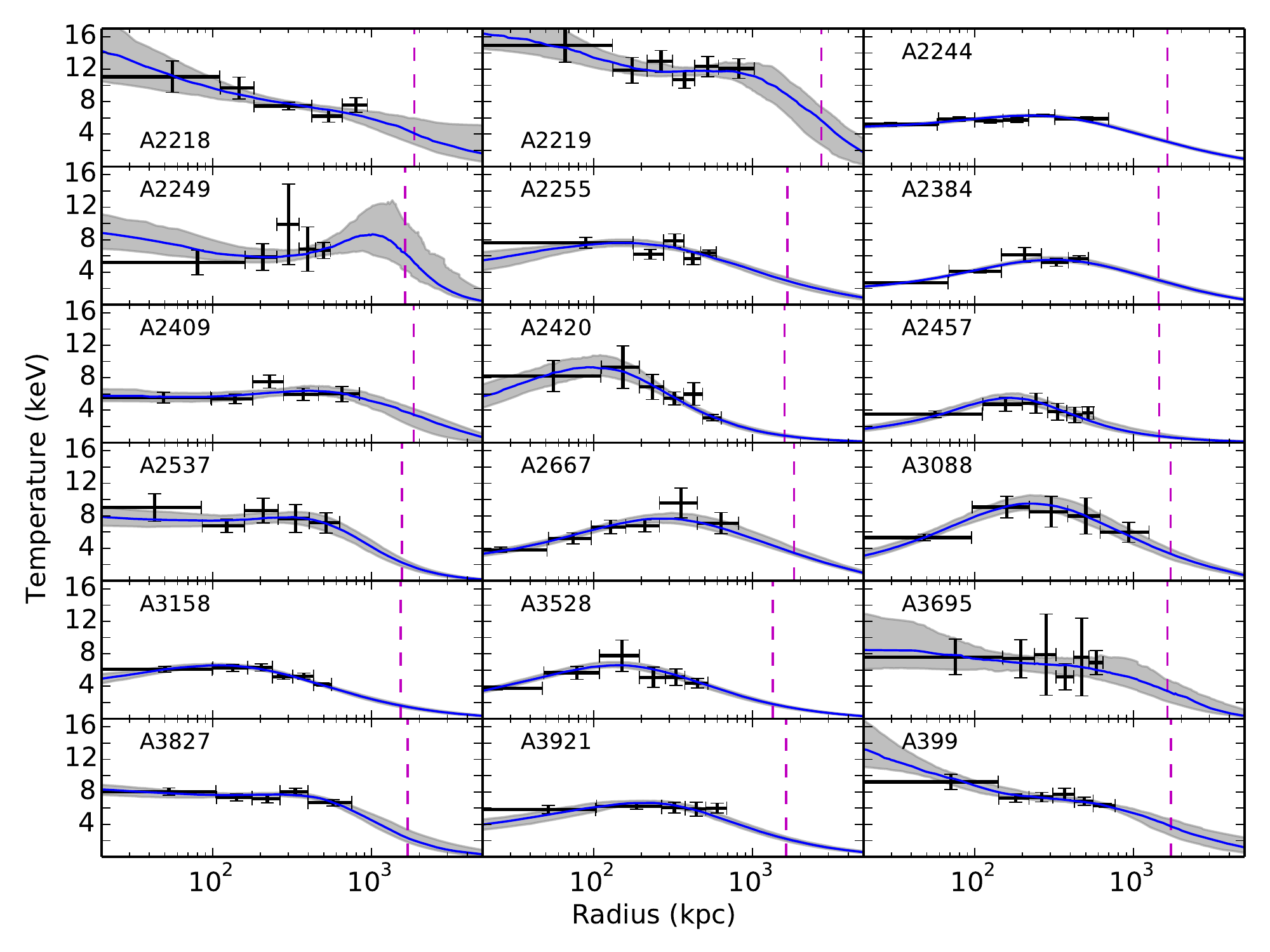}
\caption{Fitting Result Continued}

\end{figure}
\setcounter{figure}{0}
\begin{figure}

\includegraphics[width=0.95\textwidth,height=1.0\textwidth]{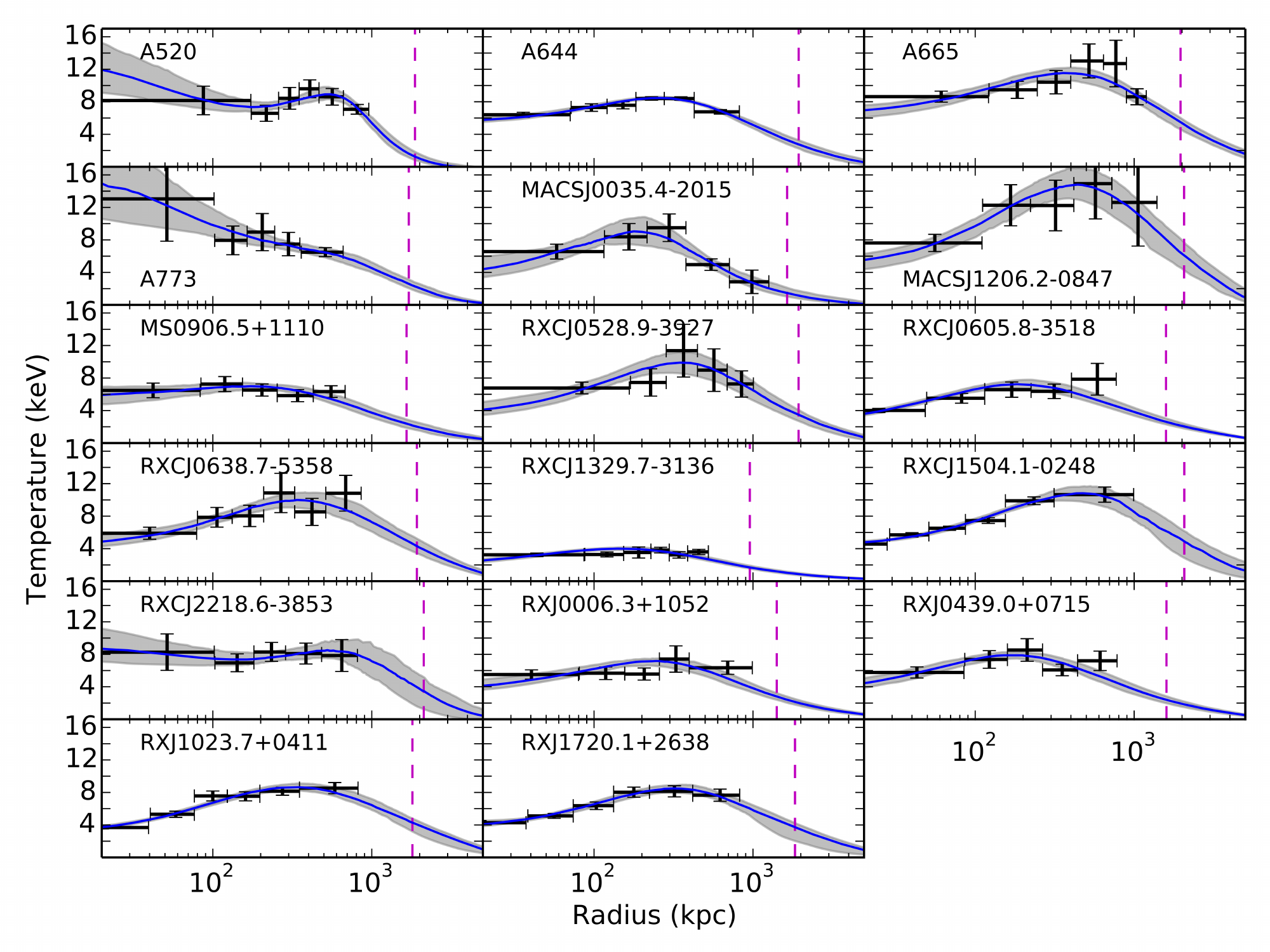}
\caption{Fitting Result Continued}

\end{figure}

\begin{figure}
\includegraphics[width=1.0\textwidth,height=0.9\textwidth]{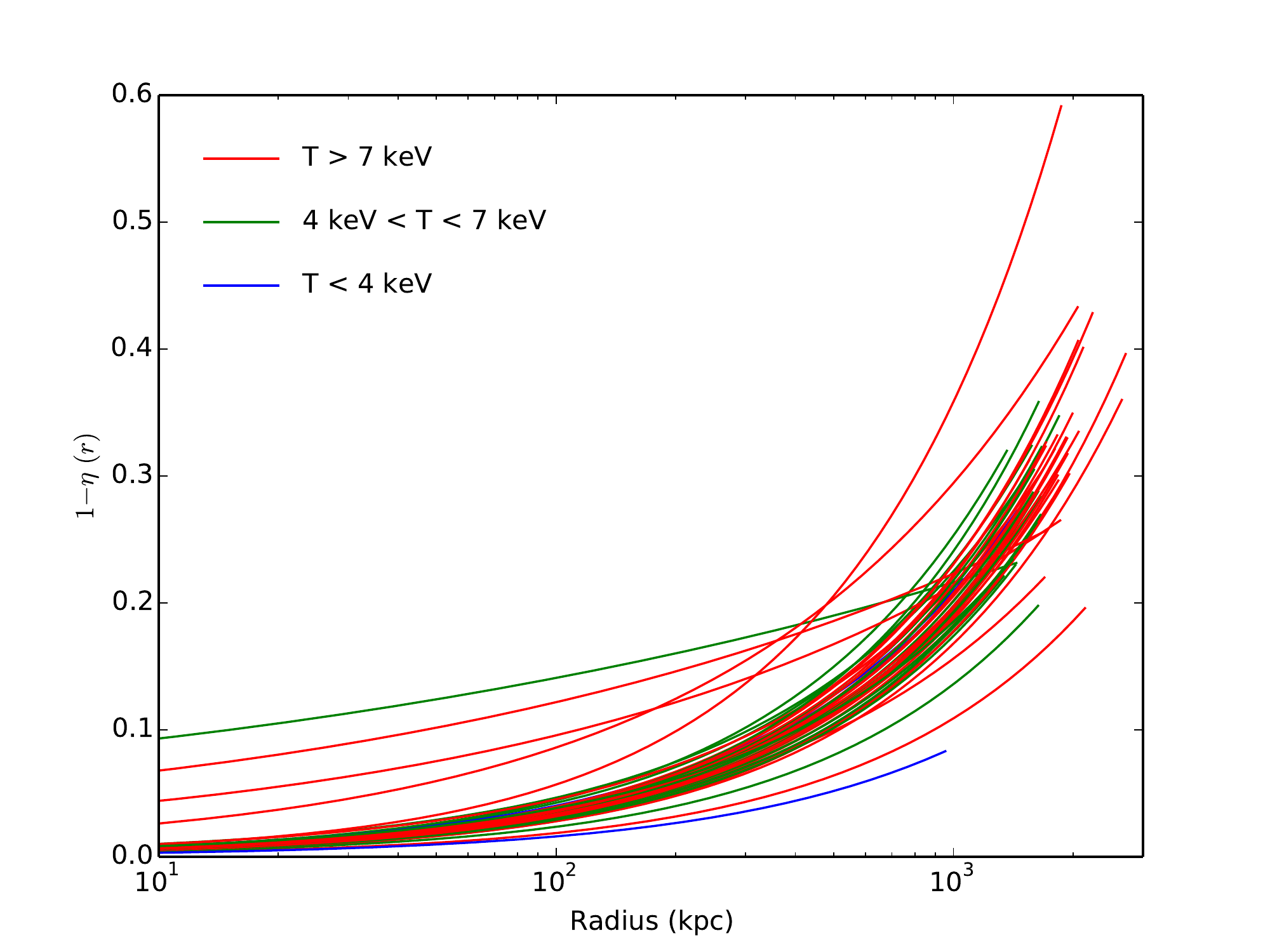}
\caption{{{Non-thermal energy fraction $1-\eta(r)$ (c.f., Eq. \ref{eta}) as an index of the violation to the hydrostatic equilibrium, which are plotted out to $1.5r_{500}$, the radius cut used to compare between \Chandra{} and \Suzaku{} results. $T$ in the legend stands for average temperature (Table \ref{sample_info}) of the cluster.}} }
\end{figure}

\begin{figure}

\includegraphics[width=1.0\textwidth,height=0.9\textwidth]{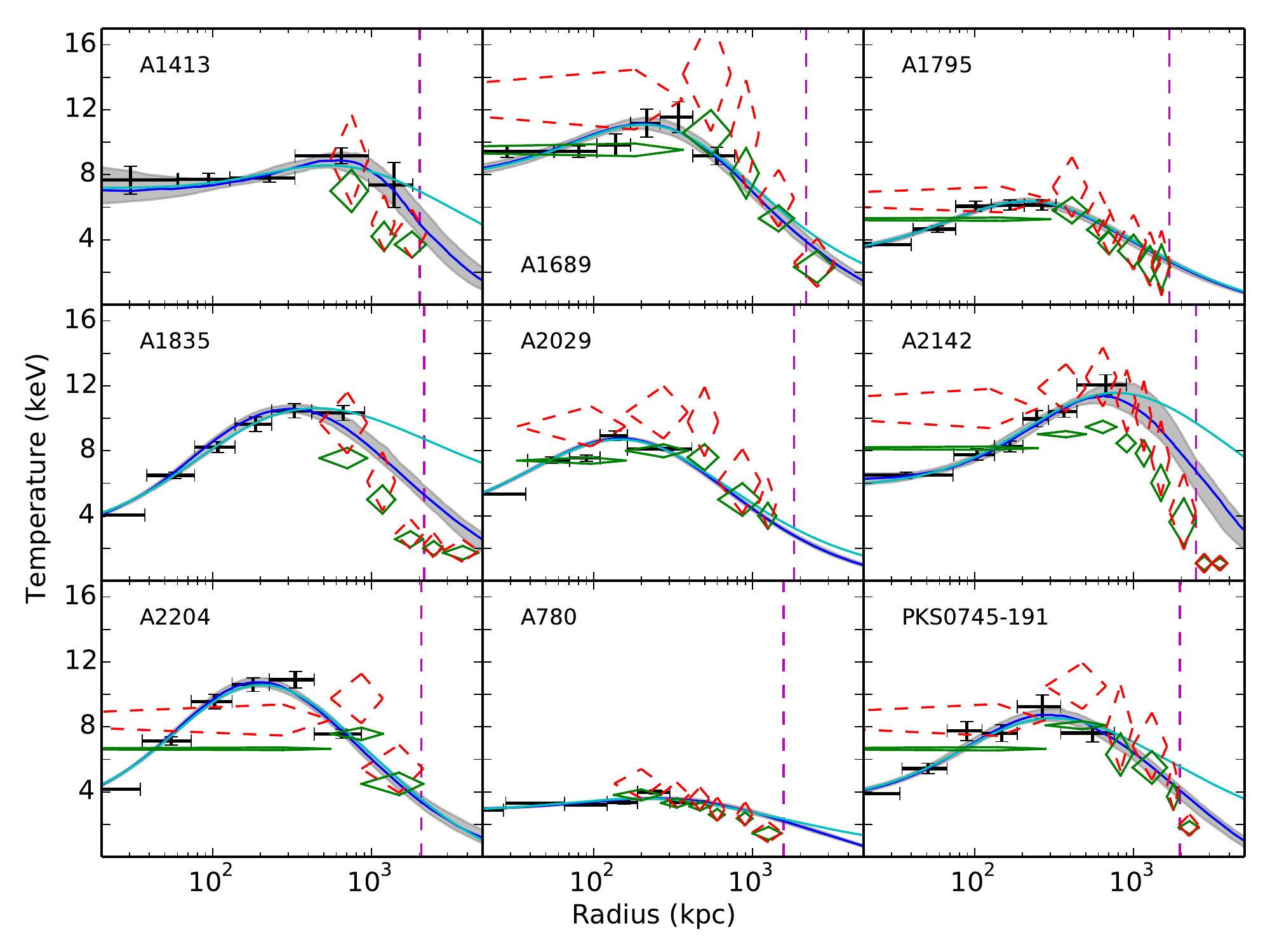}
\caption{A comparison between the predictions of our model (dark blue curves), which are obtained based on the \Chandra\ observation (dark crosses) along with the 68\% errors (shadow), and the \Suzaku\ measurements (original data are marked with green diamonds, and the data obtained after the differences between \Suzaku\ and \Chandra\ measurements are corrected are marked with red diamonds) out to $1.5r_{500}${, which} are marked with dashed lines in magenta. The light blue curves show the model predictions when the assumption of hydrostatic equilibrium is forced throughout the cluster.}

\end{figure}

\end{document}